\documentclass[aps,12pt,floatfix,tightenlines,showkeys,showpacs]{revtex4}
\usepackage[dvips]{color,graphicx}
\usepackage{graphicx}
\usepackage{amssymb}
\newcommand{\nn}{\nonumber\\}

\newcommand{\ben}{\begin{displaymath}}
\newcommand{\een}{\end{displaymath}}
\newcommand{\be}{\begin{equation}}
\newcommand{\ee}{\end{equation}}
\newcommand{\bea}{\begin{eqnarray}}
\newcommand{\eea}{\end{eqnarray}}

\newcommand{\eq}[1]{Eq.~(\ref{#1})}

\newcommand{\bfq}{{\bf q}}
                                                           
\begin{document}

\title{\bf  \hskip10cm NT@UW-08-07\\{Color Transparency in Semi-Inclusive Electroproduction of $\rho$ Mesons}}

\author{L. Frankfurt$^1$, G.A Miller$^2$, M. Strikman$^3$}

\affiliation{
$^1$  Tel Aviv University, Tel Aviv, Israel\\
$^2$ University of Washington, Seattle, WA 98195-1560\\
$^3$ Pennsylvania State University, University Park, PA 16802}

\date{\today}

\begin{abstract}

{
We  study the electroproduction of rho mesons in nuclei
at intermediate energies,
deriving a  treatment  of the energy lost 
by the rho in each step of multiple-scattering. This enables a close match between
calculations and the experimental kinematic conditions. 
 A standard Glauber calculation
is presented,  and then the effects of  color-transparency  are included.
The influence of poor experimental resolution on the extracted 
transparency is assessed.  
 The effects  of $\rho$ meson decay inside the nucleus are examined, and are typically 
about 5\% at most. 
This effect disappears rapidly
as $Q^2$ increases from about 1 to 3 GeV$^2$, causing a rise in the transparency that is not
attributable to  color transparency.
The size of color transparency effects for C and Fe nuclei is studied  
for values of $Q^2$ up to 10 GeV$^2$. 
The detailed results depend strongly on the assumed value of 
the $\rho N$ cross section.
The overall effects of color transparency are greater than about 10 \%
 for both nuclear targets if
$Q^2$ is greater than about 5 GeV$^2$. 
}
\end{abstract}

\maketitle     
 
\section{Introduction}

In the special situation of high-momentum-transfer coherent processes 
the strong interactions between hadrons and nuclei 
can be extinguished, causing  shadowing to disappear and the
nucleus to become quantum-mechanically transparent. This phenomenon is known as
color transparency \cite{ct,Frankfurt:1994hf,Miller:2007zz}. In more technical language, color transparency is the
vanishing of initial and final-state interactions, predicted by QCD to occur in
high-momentum-transfer quasi-elastic nuclear reactions. 
In these reactions, the scattering amplitudes consist of a sum of terms involving different
intermediate states and the same final state. Thus 
 one adds different contributions to obtain the scattering
amplitude. Under such conditions the effects of gluons emitted by small
color-singlet systems tend to cancel \cite{lonus} and could nearly vanish.  Thus color transparency is also known as
color coherence. 

The important dynamical question is whether or not 
small color-singlet systems, often referred to as
point-like configurations (PLC's),
are  produced as intermediate states in high momentum transfer reactions.
Perturbative QCD predicts that a PLC
is formed 
in many two-body hadronic processes
at very large momentum transfer \cite{ct,liste}. However, PLC's can also 
be formed under
non-perturbative dynamics \cite{fms1,fms2}. Therefore 
measurements of color transparency are important for clarifying the dynamics
of bound states in QCD.

Observing
color transparency requires that a PLC is formed and that the energies are
high enough so that  the PLC does not expand completely 
to the  size of a physical hadron  while
traversing the target \cite{ffs,jm,bm}. The frozen approximation must be valid.

A direct observation of high-energy color transparency in the
$ A $-dependence of diffractive di-jet production by pions was reported in \cite{dannyref}.
The results were in accord with the prediction of \cite{Frankfurt:1993it}. 
See also \cite{Frankfurt:1999tq}.
Evidence for color transparency 
(small hadronic cross-sections) has been observed in other 
types of processes, also occurring at high energy:
in the A-dependence of
$J/\psi$ photoproduction \cite{e691}, 
in the $ Q^2 $-dependence of the $ t $-slope of diffractive $ \rho^0 $
production in deep inelastic muon scattering
(where $ Q^2 $ is the invariant mass of the virtual photon and
$ t $ denotes the negative square of the 
momentum transfer from the virtual photon
to the target proton),
 and in
the energy and flavor dependences of vector
meson production in $ ep $ scattering at HERA \cite{ha}.
For all of these processes the energy is high enough so that 
the produced small-size configuration does not expand significantly as it makes its way out 
of the nucleus.

At lower energies expansion effects do occur.
Experimental studies of high momentum transfer 
 processes in $ (e,e'p)$ and (p,pp) reactions
 have so far failed to produce
convincing evidence of color transparency\cite{eva,Aclander:2004zm,slac,cebaf}.
 First data on the  reaction $A(p,2p)$ at large scattering angles
 were obtained at BNL. 
They were followed  by the dedicated experiment EVA. The final results of 
EVA \cite{Aclander:2004zm}
 can be summarized as follows.
An eikonal approximation calculation  agrees with  data for  
$\mbox{p}_p$=5.9 GeV/c, and  the transparency increases significantly for momenta up to about 
$\mbox{p}_p$= 9 GeV/c. Thus it seems that 
 momenta of the incoming proton $\sim $
10 GeV are  sufficient to  significantly suppress  expansion effects. 
Therefore one can use proton projectiles 
with  energies above $\sim $10 GeV to study other aspects of the strong interaction dynamics.
But the observed drop in transparency  for values of $\mbox{p}_p$ ranging from 
11.5 to  14.2 GeV/c 
represents a problem for all current models, 
including  
\cite{Brodsky:1987xw,Ralston:1990jj,Jennings:1990ma,Jennings:1992hs,Frankfurt:1994nn}
 because of its broad range in energy.
 This suggests that  leading-power  quark-exchange mechanism for
 elastic scattering dominates only at very large energies.

It is natural to expect that it is easier to observe color transparency  for
 the interaction/production  of mesons than 
for baryons because   only two quarks have to come close together. 
A high resolution pion production experiment
 recently reported 
evidence for the onset of CT \, \cite{:2007gqa} at Jefferson Laboratory
in the process $eA\to e\pi^+ A^*$. The pion momentum was chosen to be equal to that of the 
virtual photon,  $\vec{p}_{\pi}\|\vec {q}$ to 
 minimize the importance of  elastic rescattering effects. 
The coherence length defined as the distance between the point 
where $\gamma^*$ converted to a $q\bar q$ of invariant mass $M_{q\bar q}$
and where $q\bar q$ interacts with a 
nucleon - $|l_{in}|=2q_0/(Q^2+M^2_{q\bar q})$, corresponding 
 to the longitudinal distance between the  
point  where $\gamma^*$ knocks out a $q\bar q$ pair from the nucleon and the nucleon 
center, is 
small for the kinematics of \, \cite{:2007gqa}, can take on  both positive and negative values, 
and varies weakly with $Q^2$. 
This simplifies  the 
interpretation of the $Q^2$ dependence of the transparency 
as compared with the case of small $x$ 
where  $l_{in}$ becomes comparable to the size of the nucleus.
 The experimental results  agree well with predictions of \cite{Larson:2006ge} and
 \cite{Cosyn:2007er}.

Does the observation of  color transparency in the pion experiments imply that
the effect should be observed in the electroproduction of rho mesons? Answering this
question is the aim of the present paper,  stimulated by the existence of and the expected 
imminent publication  
of a  Jlab experiment led by Hafidi\cite{ANL}. 
The nuclear targets were the deuteron, $^{12}$C and $^{56}$Fe.
They made careful measurements with various experimental cuts.
To avoid the effects of resonances the virtual photon-proton cm energy $W$ was
taken as
 $W > 2$ GeV, with a range between 2 and 3.1 GeV.
 To ensure that the reaction process was
diffractive, the momentum transfer variable $t$ was restricted by
$-0.4 \;{\rm GeV}^2\ge  - (t-t_0) \ge- 0.1 $ GeV$^2$, 
where $t_0$ is the magnitude of the minimum squared momentum transfer to the nucleon. 
The exclusive nature of the reaction
was maximized by requiring that the ratio $z$ of the energy of the observed $\rho^0$ meson to
that of the virtual photon varies between 0.9 and 1.0. 
 The variation of 
 $Q^2$  was between  0.7  and  3.2 GeV$^2$. The photon energy $\nu$
 varies from 2.2 to 4.8 GeV for $Q^2$ between  
0.7  and  1.5 GeV$^2$ and from 2.4 to 4.6 GeV for  $Q^2$ values greater than  1.5 GeV$^2$. 
These kinematics correspond to a range of conventionally defined 
coherence lengths $l_c\equiv 2\nu/(Q^2+M_\rho^2)$
between 0.4 and 0.9 fm, which are larger than $l_{in}$. The localization of the interaction
 occurs between  the  narrow range between $-l_{in}$ and $l_{in}$. Thus, we may safely
 take the start of the space time evolution of the virtual photon to occur at  the nucleon
center. 
The experimental kinematics are defined in terms
of  $l_c$ \cite{ANL}, so we   present our results as a function of $l_c$.

Sect.~\ref{sec:gen} is concerned with general issues of the reaction theory for 
 nuclear electroproduction of
vector mesons. 
The specific nature of the kinematics provides the motivation for Sect.~\ref{sec:gl}
which develops the   theory necessary to account for these kinematic restrictions. 
In particular, we derive a  treatment  of the energy lost 
by the rho in each step of multiple-scattering that also accounts
for the momentum transfer to the nucleus. This,
done using standard  Glauber theory, is the main difference between
 the present
 effort and earlier ones such as \cite{Falter:2002vr}.
We study  the  kinematics of the escaping ejectile as a function of the  order of term in 
 the 
multiple scattering series. The  generalization to include  the effects of color
transparency is presented  in Sect.~\ref{sec:ct}.
A brief summary is presented in Sect.~\ref{sec:sum}

\section{General issues}
\label{sec:gen}

\subsection{Basics}
It is worthwhile to discuss color transparency for vector meson production from
the perspective of high energies. 
The basic idea is that the incident 
virtual photon decays spontaneously into a $q\bar{q}$ pair which
then interacts strongly with the target system.
At high energies where the space-time evolution of small wave packets is slow one can introduce a notion of the cross section of scattering of a small dipole configuration (say $q\bar q$) of transverse size $d$ on the nucleon \, \cite{Frankfurt:1993it, Blaettel:1993rd} which in the leading log approximation is given by \, \cite{Frankfurt:2000jm}
\begin{equation}
\sigma(d,x_N)= {\pi^2\over 3} \alpha_s(Q^2_{eff}) d^2\left[xG_N(x,Q^2_{eff}) +2/3 xS_N(x, Q^2_{eff})\right],
\label{pdip}
\end{equation}
where $Q^2_{eff} = \lambda/d^2, \lambda= 4 \div 10\;, x=Q^2_{eff}/s$, with
$s$ the invariant energy of the dipole-nucleon system,
 and  $S$ is the  sea quark distribution  for quarks making up the dipole.  
The  difference between \eq{pdip} and the simplest two gluon exchange model \cite{Gunion:1976iy}
of this interaction is small, but grows as the energies  increases.

For 
  hard, high-energy processes 
in which  a small dipole is produced (pion diffraction into two jets) 
or the initial state is highly localized (exclusive production of mesons for large values of 
$Q^2$),
 one can prove
  factorization theorems  which allow the scattering amplitude to be represented 
as the product of the generalized parton densities,
and wave functions in the frame in which the target is at rest and the 
projectile and the  final system have high momentum\cite{Frankfurt:1993it,Frankfurt:2000jm,Brodsky:1994kf,Collins:1996fb}. The proofs require 
  the  color transparency property in pQCD,
 understood in the sense of the suppression, $\propto d^2$, of  
multiple interactions of the dipole.
Note that  the  definition of  color transparency does not simply correspond
to the nuclear amplitude being $A$ times the nucleonic amplitude because
both $G_N$ and $S_N$ may depend upon the nuclear environment. Instead, color transparency
corresponds to  the dominance of the leading twist term in the relevant scattering amplitude
 \cite{Frankfurt:1993it}.

For  vector meson production both quark $\propto S_N$  and gluon $\propto G_N$ 
exchanges are allowed. In  pion production only exchange by a $q\bar q$ pair is allowed. 
At Jlab kinematics, corresponding to $x\ge 0.2$, Eq.~(\ref{pdip})
is dominated by the quark contribution, and mechanism dominates the production of  vector 
meson.
We also note that  pion production occurs via the knockout of a $q\bar q$ pair. 
The kinematics are that of the ERBL region.

\subsection{Reaction Theory}
\begin{figure}
\unitlength1.cm
\includegraphics[width=5.0cm] {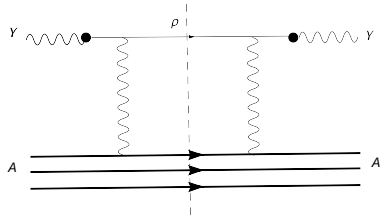} %

\caption{
Feynman diagram for the cross section. The vertical wiggly lines represent
schematically the interaction of \eq{pdip}. The portion to the right
of the dashed line represents the scattering
amplitude and the portion to the left represents the complex conjugate of the scattering
amplitude.
 }%
\label{rhofig1a0}\end{figure}

We start our analysis  of the nuclear reaction by  considering  the 
relevant Feynman diagrams for the $\rho$-meson production cross section.
 There are two classes of diagrams. One is  corresponding  to 
the quasi-classical picture  in which  
$\rho$ meson scatters from the same nucleons in both the {\it in} and {\it out} states, 
Fig.~\ref{rhofig1a0}.  We apply this picture for kinematics such that the struck 
nucleon is knocked out of the nucleus. If $-t\ge 6/R_A^2$, the square of the elastic
nuclear (of radius $R_A$) form factor will be very small, so that terms in which the nucleus remains in
its ground state are absent. 
This condition is satisfied for the Jlab kinematics.

\begin{figure}
\unitlength1.cm
\includegraphics[width=5.0cm] {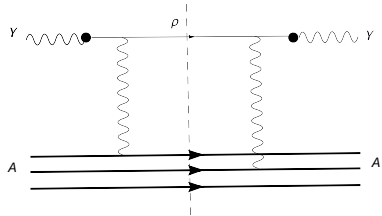} %

\caption{
 Feynman diagram for the cross section. The vertical wiggly lines represent
schematically the interaction of \eq{pdip}. The portion to the right
of the dashed line represents the scattering
amplitude and the portion to the left represents the complex conjugate of the scattering
amplitude.
 }
\label{rhofig1a}\end{figure}

\begin{figure}%
\unitlength1.cm
\includegraphics[width=5.0cm]{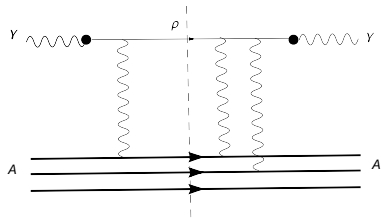} %
\includegraphics{rhofig1b.eps}
\caption{ Suppressed Feynman diagram for the cross section. The portion to the right
of the dashed line represents the scattering
amplitude and the portion to the left represents the complex conjugate of the scattering
amplitude.
 }%
\label{rhofig1b}\end{figure}

The second class includes all other diagrams 
which correspond to interference of interaction with different nucleons. The simplest two of them are shown in
Fig.~\ref{rhofig1a} and Fig.~\ref{rhofig1b}.
The simplest one (Fig.~\ref{rhofig1a}) was considered in \cite{Frankfurt:2000jm}. 
  In contrast with the diagram of Fig.~\ref{rhofig1a0}, these terms involve  the two-nucleon
correlation function, which in momentum space,  
is approximately proportional to the square of the nuclear form factor, $F_A(t)^2$. Thus these
terms are substantially suppressed.
For the region of $t$ relevant for the Jlab experiments 
one can safely neglect the diagrams of the second kind.

Therefore  we may organize the calculation  in the following way.
The production of the $\rho$ takes place on a given nucleon, and rescattering
on other nucleons are expected to vanish if color transparency holds.
 
The transparency $T_A$ is  defined here as the ratio of the observed cross section
to $A$ (the nucleon number) times the cross section on a free nucleon, with
perfect transparency occurring for $T_A\rightarrow1$.
Thus we start with the notation for the 
 the rho meson production cross section on a single
nucleon, ${d\sigma^{\gamma*}\over dt}$:
\bea
{d\sigma^{\gamma*}\over dt}=\left({d\sigma^{\gamma*}\over dt}\right)_{t=t_0}\exp{(-B_1 q^2)},\eea
where the transverse momentum transfer variable
$q^2\equiv \bfq^2=-t$ ($\bfq$ is in the $\perp$ direction)
 and $t_0$ corresponds to  the minimum  momentum transfer to the nucleon.
The produced $\rho$ then can interact with nucleons as it makes its way out of
the nucleus. The   $\rho-$nucleon 
scattering cross section is given by 
${d\sigma^{V}\over dt}$ with
\bea
{d\sigma^{V}\over dt}=\left({d\sigma^{V}\over dt}\right)_{t=0}\exp{(-B_2 q_\perp^2)}.\eea
in the t-range relevant for the JLab CT experiment.
The total vector meson-nucleon cross section is denoted as 
 $\sigma_{\rm tot}.$  It is traditional \cite{Bauer:1977iq} to
assume that this cross section is the same as that for pions.
One could use the 
 Particle Data Group  parameterization \cite{pdg} for the pion cross sections. However, 
the accuracy of this assumption has not been quantitatively tested, so
we shall report results using
a small range of values of   $\sigma_{\rm tot}$ between 25 and 30 mb. 
A convenient
parametrization of the vector meson-nucleon scattering cross section, based on unitarity is:
\bea
{d\sigma^{V}\over dt}={\sigma_{\rm tot}^2\over 16\pi}(1+\alpha^2)\exp{(-B_2 q_\perp^2)},
\label{unitary}\eea
where $\alpha$ is the ratio of the real to imaginary parts of the  elastic scattering amplitude.
Values of the parameters are given in Table I.

The nuclear rho production cross section  ${d\sigma\over dt}$ results from the production of a $\rho$ meson by 
by a single nucleon, followed by elastic rescattering on the surrounding nucleons. We thus
express ${d\sigma\over dt}$ as
\bea
{d\sigma\over dt}=\sum_{n=0}^\infty {d\sigma_n\over dt},\eea
where $n$ denotes the number of elastic rescattering terms. We  explicitly count
the number of elastic rescattering terms    to  keep track of the energy loss and 
ensure that we match the experimental conditions. We assume that any contribution to the
vector meson-nucleon {\it inelastic} cross section 
involves large enough energy loss to be cut out
of the ANL experiment.

We define the transparency  as
\bea T_n\equiv\frac{{d\sigma_n\over dt}}{A {d\sigma^{\gamma^{*}}\over dt}}\\
T_A=\frac{{d\sigma_n\over dt}}{A {d\sigma^{\gamma^{*}}\over dt}}=\sum_{n=0}^\infty T_n.\label{tndef}\eea

\begin{table} 
\caption{Values of parameters used in the calculation.\\}
 
\begin{tabular}{|c|c|c|c|c|c|}\hline
$\left({d\sigma^{\gamma*}\over dt}\right)_{t=t_0}$  & $B_1$ & 
$\sigma_{\rm tot}$ &  $B_2=B_{\rm soft}$ & $B_{\rm hard}$ & $R,a$  \\
 mb GeV$^{-2}$ & $GeV^{-2}$&mb &$GeV^{-2}$ &$GeV^{-2}$ &fm \\ \hline
1.25 $-$ 1.5 & 6     & 25  & 6  &2-4 
&1.1,0.54 \\\hline
\end{tabular}
\label{alpha}
\end{table}

\section{Glauber Formulae}
\label{sec:gl}
In the absence of the effects of color transparency,  one expects that 
Glauber theory would provide a reasonable description of the data. Thus this 
theory serves as   our starting point. 

If no elastic rescattering takes place, the cross section is given by
\bea
&&{d\sigma_0\over dt}=A{d\sigma^{\gamma*}\over dt}\int\;d^2b\;\int^\infty_{-\infty}dz\; \rho(b,z)
\left(1-\sigma_{\rm tot}T(b,z)\right)^{A-1},\\
&&T(b,z)\equiv \int_z^\infty \;dz'\;\rho(b,z').\label{zero}\eea
Here  the  nuclear density $\rho(r)=\rho(b,z)=\rho(\sqrt{b^2+z^2})$ is normalized to unity.
We take for heavy nuclei
\bea \rho(r)={\rho_0\over 1+e^{r-R\over a}},\eea
with $R=1.1 A^{1/3}$ fm, and $a$=0.54 fm.
Combining \eq{zero} with \eq{tndef} yields
\bea T_0=\int\;d^2b\;\int^\infty_{-\infty}dz\; \rho(b,z)
\left(1-\sigma_{\rm tot}T(b,z)\right)^{A-1}.\label{T0}\eea
If a single elastic rescattering of the rho meson occurs one obtains a contribution,
${d\sigma_1\over dt}$ with 
\bea
&&{d\sigma_1\over dt}= 
A(A-1)\left({d\sigma^{\gamma*}\over dt}\right)_{t=t_0}\left({d\sigma^{V}\over dt}\right)_{t=t_0}
\int\;d^2b\;\int^\infty_{-\infty}dz 
\rho(b,z)T(b,z)
\left(1-\sigma_{\rm tot}T(b,z)\right)^{A-2}\nn
&&\int 
{d^2q_1\over\pi}
{d^2q_2\over\pi}e^{-B_1q_1^2-B_2q_2^2}\delta^{(2)}
(\bfq_1+\bfq_2-\bfq)
.\label{one}\eea
The integral over $q_1,q_2$ can be evaluated with the
result
\bea
&\int 
{d^2q_1\over\pi}
{d^2q_2\over\pi}e^{-B_1q_1^2-B_2q_2^2}\delta^{(2)}
(\bfq_1+\bfq_2-\bfq)={1\over\pi(B_1+B_2)}\exp{(-{B_1B_2\over B_1+B_2}q^2)}.\eea
There is a hundred MeV cutoff on the nuclear excitation energy. 
The single scattering term above leads to nuclear excitation
energies less than that, so the  
cutoff is not effective in this term and
does not enter here. 
 Then 
\bea T_1=(A-1){1\over\pi(B_1+B_2)}\exp{({B_1^2\over B_1+B_2}q^2)}{\sigma_{\rm tot}^2\over 16\pi}(1+\alpha^2)\nn
\int\;d^2b\;\int^\infty_{-\infty}dz 
\rho(b,z)T(b,z)
\left(1-\sigma_{\rm tot}T(b,z)\right)^{A-2}.\label{t1}\eea

The double rescattering term ${d\sigma_2\over dt}$ is given by 
\bea&&{d\sigma_2\over dt}={A(A-1)(A-2)} 
\left({d\sigma^{\gamma*}\over dt}\right)_{t=t_0}
\left[\left({d\sigma^{V}\over dt}\right)_{t=t_0}\right]^2\nonumber\\&&
\int\;d^2
b\;\int^\infty_{-\infty}dz \rho(b,z){1\over2}\int_z^{\infty} dz'\rho(b,z')\int_{z}^\infty  dz_2\rho(b,z_2)\nn&&
\left(1-\sigma_{\rm tot}T(b,z)\right)^{A-3}\int {d^2q_2\over\pi}\int{d^2q_3\over\pi}
e^{-B_1q_1^2-B_2q_2^2-B_2q_3^2}\delta^{(2)}
(\bfq_1+\bfq_2+\bfq_3-\bfq)\label{comp}
.\eea
This term can be simplified to 
\bea
&&{d\sigma_2\over dt}={A(A-1)(A-2)\over2}\left({d\sigma^{\gamma*}\over dt}\right)_{t=t_0}
\left[\left({d\sigma^{V}\over dt}\right)_{t=t_0}\right]^2
\nn&&
\int\;d^2b\;\int^\infty_{-\infty}dz \rho(b,z)
T^2(b,z)\left(1-\sigma_{\rm tot}T(b,z)\right)^{A-3}\nn&&\int
{d^2q_1\over\pi}
 {d^2q_2\over\pi}\int{d^2q_3\over\pi}
e^{-B_1q_1^2-B_2q_2^2-B_2q_3^2}\delta^{(2)}
(\bfq_1+\bfq_2+\bfq_3-\bfq)
.\label{simp}\eea
The result \eq{simp} is useful for the Glauber calculations, but \eq{comp} is more easily generalized to include the effects of color transparency.
The integral over $q_1,q_2,q_3$ can be done with the result
\bea &&
T_2={2\over \pi B_2(4B_1+B_2)}\exp{\left({4B_1B_2 q^2\over(4B_1+B_2)}\right)}
{(A-1)(A-2)\over2}
\left[\left({d\sigma^{V}\over dt}\right)_{t=t_0}\right]^2
\nn&&
\int\;d^2b\;\int^\infty_{-\infty}dz \rho(b,z)
T^2(b,z)\left(1-\sigma_{\rm tot}T(b,z)\right)^{A-3}.\label{t2}\eea

\begin{figure}%
\unitlength1.cm
\includegraphics[width=7cm]{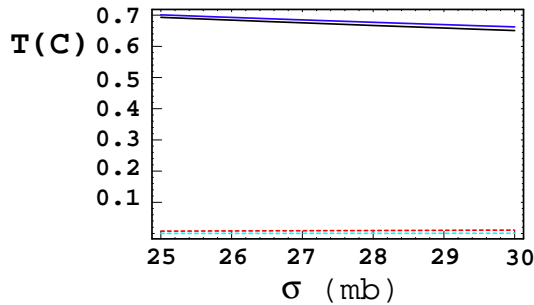}
\caption{(Color online) Glauber calculations for $\rho $ production on $C$.
 Transparency {\it vs.} $\sigma_{\rm tot}=\sigma$. The heavy solid (black)
 curve represents $T_0$, the long-dashed (red) curve $T_1$, and the short dashed  
(cyan) curve $T_2$. The sum $T_0+T_1+T_2$ is shown an the solid thin (blue) curve. The forward limit,
no transverse momentum transfer is used.
 }%
\label{figl}\end{figure}

\begin{figure}%
\unitlength1.cm
\includegraphics[width=7cm]{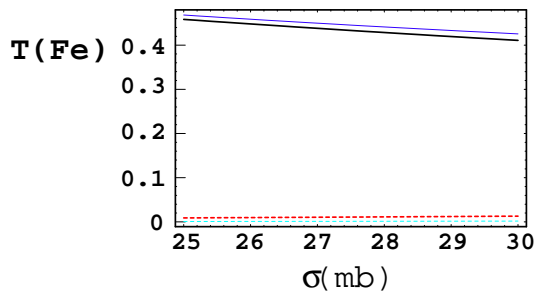}
\caption{(Color online) Glauber calculations for $\rho $ production on $Fe$.
 Transparency {\it vs.} $\sigma_{\rm tot}=\sigma$. The heavy solid (black)
 curve represents $T_0$, the long-dashed (red) curve $T_1$, and the short dashed  
(cyan) curve $T_2$. The sum $T_0+T_1+T_2$ is shown an the solid thin (blue) curve. The forward limit,
no transverse momentum transfer is used.
 }%
\label{fig2}\end{figure}

Numerical results for the C and Fe targets
are presented in Figs.~\ref{figl} and ~\ref{fig2}. The effects of $T_0$ are 
dominant. The limit of $q^2=0$ is used here in evaluating \eq{t1} and \eq{t2}.
Using $q^2$ = 0.1 GeV$^2$, (the value for which all contributions are largest)
  leads (for the$Fe$ target) to enhancements of a 1.3 for $T_1$ and $1.6$ for $T_2$.
These are not sufficient to modify the conclusion that the effects of $T_0$ are dominant.
These effects are much smaller for the carbon nucleus. Therefore we shall only be concerned
with $T_0$ in future sections on color transparency. 

\subsection{Dependence on momentum transferred to the nucleus}
\begin{figure}
\unitlength1.cm
\includegraphics[width=7cm]{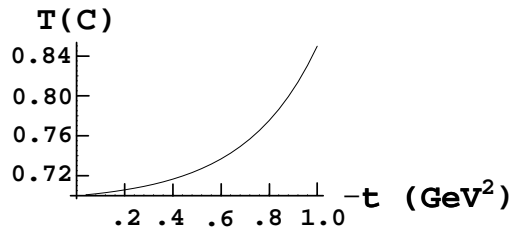}
\caption{ Glauber calculations for $\rho $ production on $C$.
 Transparency, $T_0+T_1+T_2$, {\it vs.} $-(t-t_0)$, $\sigma_{\rm tot}=25$ mb. 
 }%
\label{c12qsqdep}\end{figure}

\begin{figure}%
\unitlength1.cm
\includegraphics[width=7cm]{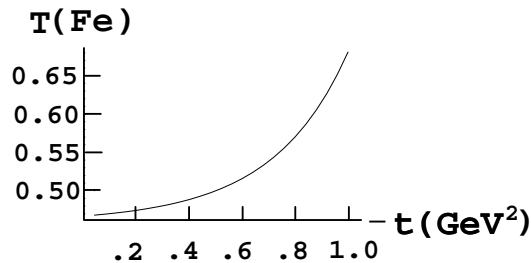}
\caption{ Glauber calculations for $\rho $ production on $Fe$.
 Transparency, $T_0+T_1+T_2$, {\it vs.}$-(t-t_0)$, $\sigma_{\rm tot}=25    $ mb. 
 }
\label{fetdep}\end{figure}

So far we have taken $\bfq^2=0$. 
This factor appears in the equations \eq{t1}, \eq{t2} for $T_1,T_2$.
However, taking $\bfq^2\ne0$ enhances both of these corrections to the leading term $T_0$.
To illustrate the importance of controlling the kinematics, we display
the total transparency, $T=_0+T_1+T_2$. for $^{12}$C and $^{56}$Fe as a function of
$t$ in Figs.~\ref{c12qsqdep} and \ref{fetdep}. One sees a rapid rise of $T$ as 
$-t$ increases. Experiments of resolution  that effectively integrates
 over $t$,  therefore observe a mix of transparencies. A change of acceptance in
$t$ that varies with the photon virtuality, $Q^2$,  
could mimic the turning on of color transparency. The theory of \cite{Kopeliovich:2001xj}
presents $t$-integrated values of the transparency, obtained using a  dipole expression
for the $q\bar q$ nucleon cross section that describes the DGLAP region quite well. 


\subsection{Glauber Treatment of Rho Decay}
We explain how the effects of $\rho$ decay to $\pi\pi$ are to be included in the
calculations. This is to guide experimentalists in making the necessary corrections.
The results of the previous sub-section  are to be compared with experimental data which
have been corrected for the effect. If the data have not been subject to this
correction, this effect should be included.

The influence of $\rho$ decay
 is implemented through the replacement
\bea  \sigma_{\rm tot}T(b,z)\rightarrow       \int_z^\infty dz'\rho(b,z')\widehat{\sigma}(z'-z),\eea
where 
\bea \widehat{\sigma}(z'-z)=\sigma_{\rm tot}\exp{[-{\Gamma m_\rho\over \sqrt{\nu^2-m_\rho^2}}\;(z'-z)]}+2\sigma_{\pi N}
\left(1-\exp{[-{\Gamma m_\rho\over \sqrt{\nu^2-m_\rho^2}}\;(z'-z)]}\right).\eea
The interpretation of this is that the produced vector meson state has two components, $\rho$ and $2\pi$
with a total probability of unity. As time goes by and $z'$ increases from $z$, the $\rho$ component
decays away and the two pion component grows. In the limit that $\nu\rightarrow\infty$, one
finds $\sigma_{\rm eff}(z'-z))\rightarrow \sigma_{\rm tot},$ which is the desired limit.

In the following, we make a simplified estimate using
\bea
\sigma_{\pi N}=\sigma_{\rm tot},\eea so that
\bea 
\sigma_{\rm eff}(z'-z)=\sigma_{\rm tot}+\sigma_{\rm tot}\left(1-\exp{[-{\Gamma m_\rho\over \sqrt{\nu^2-m_\rho^2}}\;(z'-z)]}\right).
\eea
Again one can see that in the $\nu\to\infty$ limit, the correction term vanishes.
The effects of decay are to replace $T(b,z)$ of \eq{T0} with $T_D(b,z)$, with
\bea
 T_D(b,z)=T(b,z)+\int_z^\infty\;dz'\rho(b,z')\left(1-\exp{[-{\Gamma m_\rho\over \sqrt{\nu^2-m_\rho^2}}\;(z'-z)]}\right)\eea

\begin{figure}
\unitlength1.cm
\begin{picture}(14,8.2)(2.,-.6)
\includegraphics{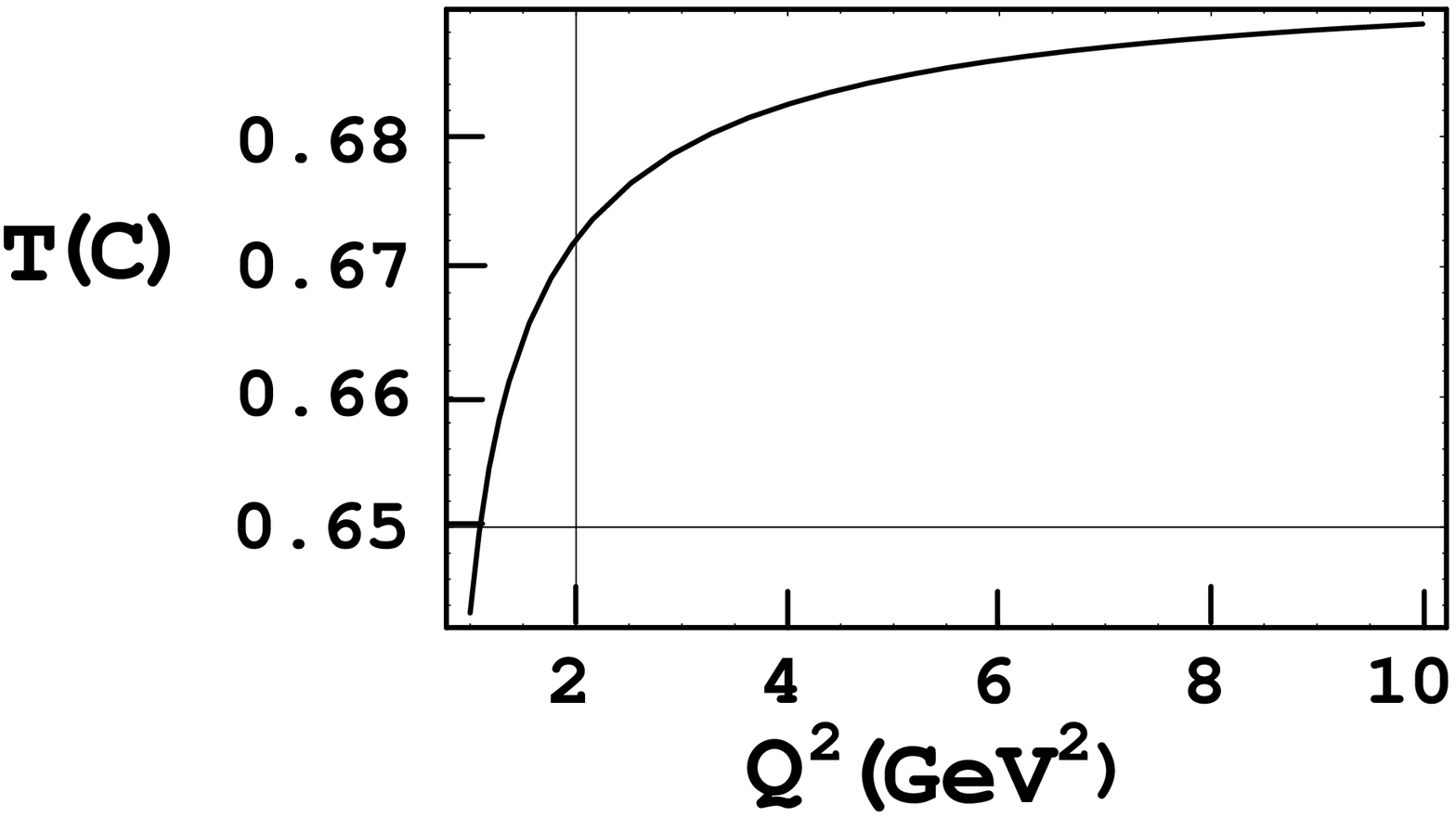}
\end{picture}
\caption{ Glauber calculations for $\rho $ production on $^{12}C$, including the 
effects of rho decay on $T_0$.
 Transparency {\it vs.} $\rho$ meson energy, $\nu$. }
\label{figcgldecay}\end{figure}

\begin{figure}
\unitlength1.cm
\begin{picture}(14,8.2)(2.,-.6)
\includegraphics{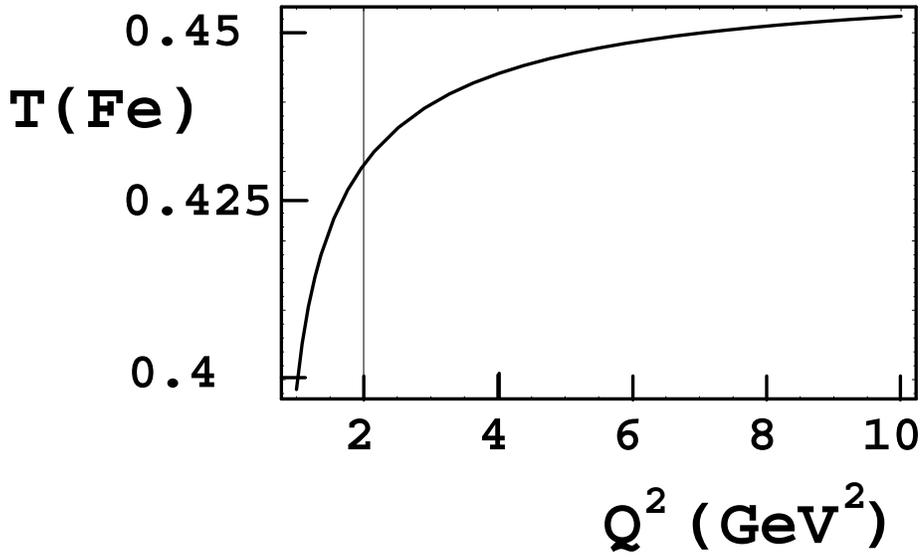}
\end{picture}
\caption{ Glauber calculations for $\rho $ production on $^{56}Fe$, including the 
effects of rho decay on $T_0$.
 Transparency {\it vs.} $\rho$ meson energy, $\nu$. 
 }
\label{feglauberdecay}\end{figure}

Numerical results Figs.~\ref{figcgldecay} and \ref{feglauberdecay} 
show that this effect is about 5\% at low $\rho$
meson energies and much smaller at higher energies. The 5\% rise should not be interpretated
as being related to the onset of color transparency. 

\section{Color Transparency}
\label{sec:ct}

This section is concerned with implementing the effects of  color transparency. In the ideal
situation of  full color transparency one would simply neglect the effects of 
final state interactions  
and then one would find $T_0$=1 under the assumption that the nuclear dependence of
$G_N,S_N$ may be neglected. However, for presently realistic experimental kinematics,
the $q\bar{q}$ pair is  produced with  a small,  non-zero size, and expands as it moves
through the nucleus.  Therefore the effects of rescattering are expected to be considerable.
Indeed, performing the experiment at increasing values of $Q^2$ leads to producing
$\rho$ mesons of increasing energy, turning off effects of expansion and the related final
state interactions. Observing this  would amount to observing the onset of color transparency.

The physics of the expansion needs to be incorporated in realistic calculations. 
See the reviews \cite{Miller:2007zz,fs,fms1,Frankfurt:1994hf}.  One technique
is to express the putative PLC as a coherent superposition of hadronic states, such as in 
 Jennings \& Miller \cite{jm}. One may also use
 the quark-based treatment of  Frankfurt \& Strikman \cite{Farrar:1988me}.
We use the latter which is more convenient, but the results of the two different
formalisms are very similar \cite{Frankfurt:1994hf}.
The results \eq{zero},\eq{one} and \eq{comp} need to be modified. In those expressions
the total cross section is replaced by an effective cross section,
$\sigma_{eff}$, which  takes the changing size of the ejectile into account.
  The effective interaction
 contains two parts, one for a propagation distance $z'-z$  less than a length $l_h$ 
describing the interaction 
of the expanding PLC, 
another, for larger values of $z'-z$
describing the final state
 interaction of the physical particle.  We use the expression \cite{Farrar:1988me}
\bea \sigma_{\rm eff}
(z, p_{\rho}) = \sigma_{\rm tot}(p_{\rho})
 \left[\left(\frac{n^2\langle  k_t^2 \rangle}{Q^2} +
 \frac{z}{l_h}(1 - \frac{n^2 \langle  k_t^2 \rangle}{Q^2} \right)\theta(l_h-z) +\theta(z-l_h)\right],
\label{sigplc}  \eea
where $l_h = 2 p_{\rho}/ \Delta M^2$, with $\Delta M^2=0.7{\rm GeV}^2$.
The prediction that the interaction of the PLC will be approximately proportional to
the propagation distance $z$ for $z<l_h$ is called the quantum diffusion model.
$n^2\langle  k_t^2 \rangle / Q^2$. In \eq{sigplc} the term 
\bea\sigma_{PLC}\equiv \sigma_{\rm tot}(p_{\rho})\frac{n^2\langle  k_t^2 \rangle}{Q^2}\label{plcp}\eea 
is the cross section for the
initially-produced PLC with 
 $n = 2$ 
and $\langle  k_t^2 \rangle^{1/2} \simeq$0.35 GeV.   

The coherence length, $l_h$ sets the time scale for the PLC to evolve and 
determines the probability that a particle experiences reduced
 PLC interactions before leaving the nuclear matter.  For propagation 
distances $z> l_h$ 
the PLC interaction is that of a  standard final 
state interaction with $\sigma_{\rm eff} \simeq \sigma_{\rm tot}(p_{\rho})$, and is 
that of a typical  Glauber-like calculation. In the limit 
 $l_h=0$ a PLC is not created 
and the calculation reduces to a Glauber-like calculation.

We also  include the effects of $\rho$ decay into $\pi\pi$
in the medium. These also modify color transparency as well as the Glauber calculation.
Suppose the $\rho$ has a width $\Gamma =(149\;{\rm MeV})$ \cite{pdg}, and has a 
Lorentz factor $\gamma=E_\rho/m_\rho$.
Then, assuming the two pions have approximately equal momenta,
 \eq{sigplc} is modified to
\bea &&\sigma^D_{\rm eff}
(z, p_{\rho}) = \sigma_{\rm tot}(p_{\rho})
 \left[\left(\frac{n^2\langle  k_t^2 \rangle}{Q^2} +
 \frac{z}{l_h}(1 - \frac{n^2 \langle  k_t^2 \rangle}{Q^2} \right)\theta(l_h-z) +\theta(z-l_h)\exp{({-\Gamma\over\gamma}z)}\right]\nn&&
+
  2\sigma_{\rm tot}(\pi N)(p_{\rho}/2)
 \theta(z-l_h)
(1-\exp{({-\Gamma\over\gamma}z)}),
\label{sigplcdec}  \eea
where $n=2$. %

The effects of CT change the result \eq{T0} to
\bea T_0=\int\;d^2b\;\int^\infty_{-\infty}dz\; \rho(b,z)
\left(1-\sigma_{\rm tot}T_{CT}(b,z,\nu,Q^2)\right)^{A-1},\label{T0CT}\eea
with 
\bea &&T_{CT}(b,z,\nu,Q^2)=\int_z^\infty dz'\;\rho(b,z'){\sigma_{\rm eff}(z'-z, p_{\rho})/ \sigma_{\rm tot}}
%
\eea
or 
\bea T_0^D=\int\;d^2b\;\int^\infty_{-\infty}dz\; \rho(b,z)
\left(1-\sigma_{\rm tot}T_{CT}^D(b,z,\nu,Q^2)\right)^{A-1},\label{T0CTD}\eea
with 
\bea &&T_{CT}^D(b,z,\nu,Q^2)=\int_z^\infty dz'\;\rho(b,z'){\sigma^D_{\rm eff}(z'-z, p_{\rho})/ \sigma_{\rm tot}}\eea

The elastic scattering cross section could be 
 substantially modified when the effects of color transparency are
included. The expression \eq{unitary} becomes
\bea
{d\sigma^{V}_{CT}\over dt}
={\sigma_{\rm eff}(z,p_\rho)^2\over 16\pi}(1+\alpha^2)\exp{(-B_2^{\rm CT} q_\perp^2)},
\label{unitaryCT}\eea
where $B_2^{\rm CT}$ depends on the propagation length $z$ as:
\bea B_2^{\rm CT}(z)=(B_{\rm hard} + \left(B_{\rm soft}-B_{\rm hard}\right) {z\over l_c}
\Theta (l_c-z)  
+ B_{\rm soft}\Theta (-l_{c}+z)\eea
This is based on the assumption that $B_2$ is a radius squared that comes from the nucleon
and the rho. The nucleon moves with low momentum here so its size  is not modified by the effects color transparency, but 
the $\rho$ is produced at high momentum transfer. Its effective radius is small (before the
influence of expansion) but increases as the PLC moves.

In our calculations we shall be concerned with the leading terms of \eq{T0CT} and \eq{T0CTD},
taking the single nucleon $\rho$ production cross 
section to be independent of the nuclear environment. The effects of \eq{unitaryCT} enter
into the higher order terms that are negligible.

\subsection{Results}
\begin{figure}
\unitlength1.cm
\begin{picture}(14,8.2)(1.5,-.6)
\includegraphics{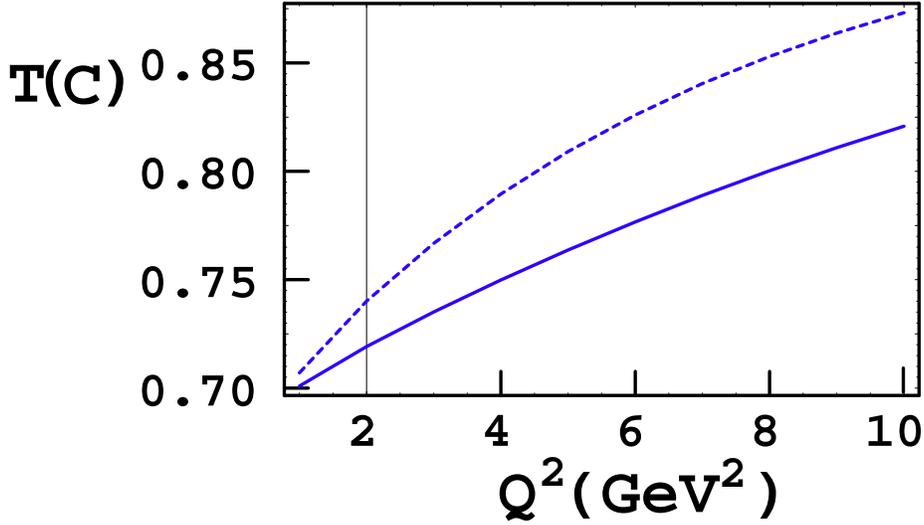}
\end{picture}
\caption{(Color online) $^{12}$C  Color transparency as a function of $Q^2$
using $T_0$ of \eq{T0CT} and \eq{sigplc} with $\sigma_{\rm tot}=25$  mb. The upper dashed  curve is computed using $l_c={2\nu\over (M_\rho^2+Q^2)} $ 
= 0.85 fm. The lower solid  curve is computed using $l_c= $ 
 0.45 fm.   }
\label{c12ct25}\end{figure}

\begin{figure}
\unitlength1.cm
\begin{picture}(14,8.2)(1.5,-.6)
\includegraphics{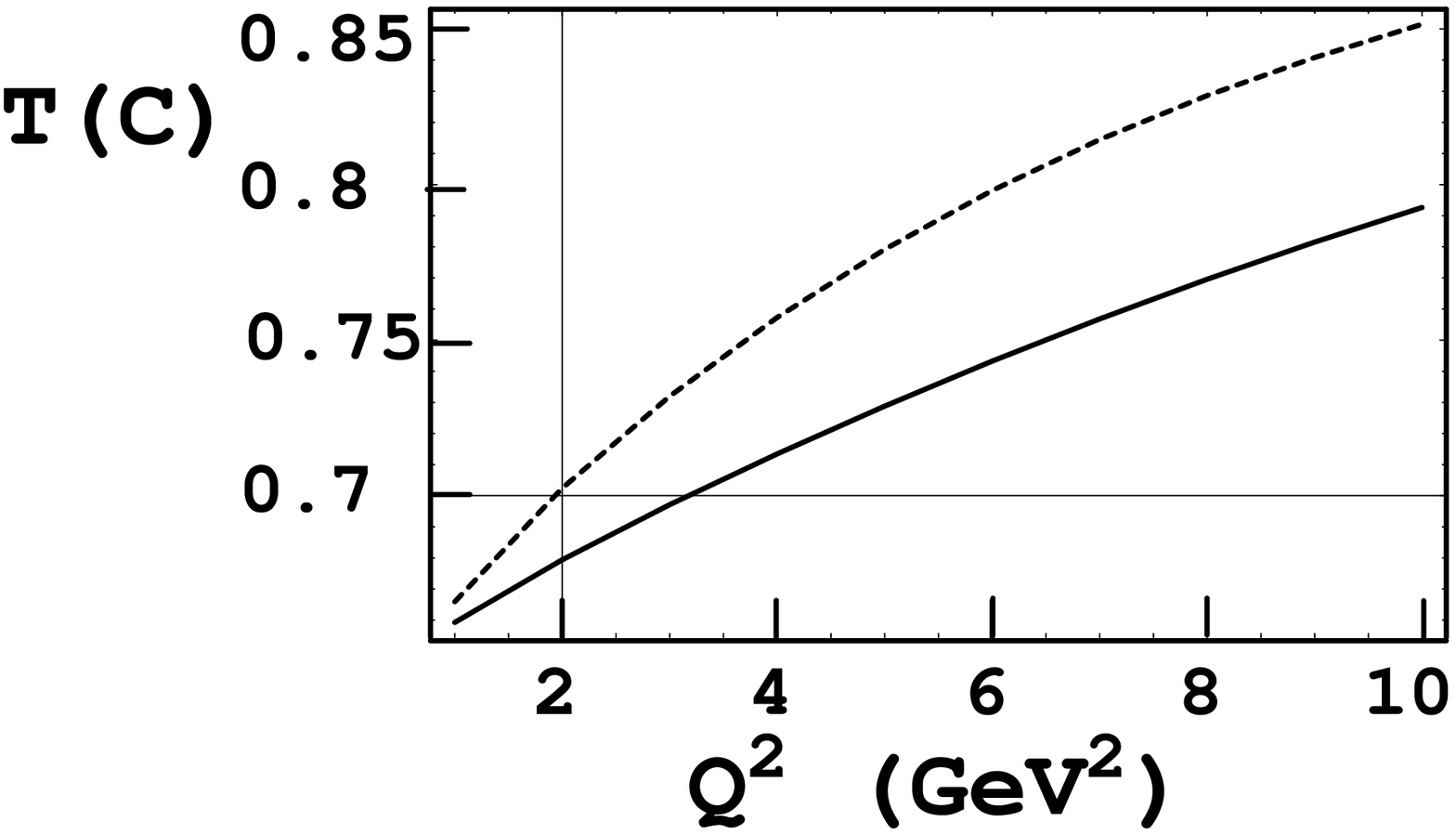}
\end{picture}
\caption{$^{12}$C Color transparency according to  \eq{T0CT} and \eq{sigplc}. Here
 $\sigma_{\rm tot}=30$  mb. The upper dashed  curve is computed using
  $l_c={2\nu\over (M_\rho^2+Q^2)} $ 
= 0.85 fm.  The lower solid   curve is computed using $l_c= $ 
 0.45 fm. }%
\label{c12ct3}\end{figure}

\begin{figure}
\unitlength1.cm
\begin{picture}(14,8.2)(1.5,-.6)
\includegraphics{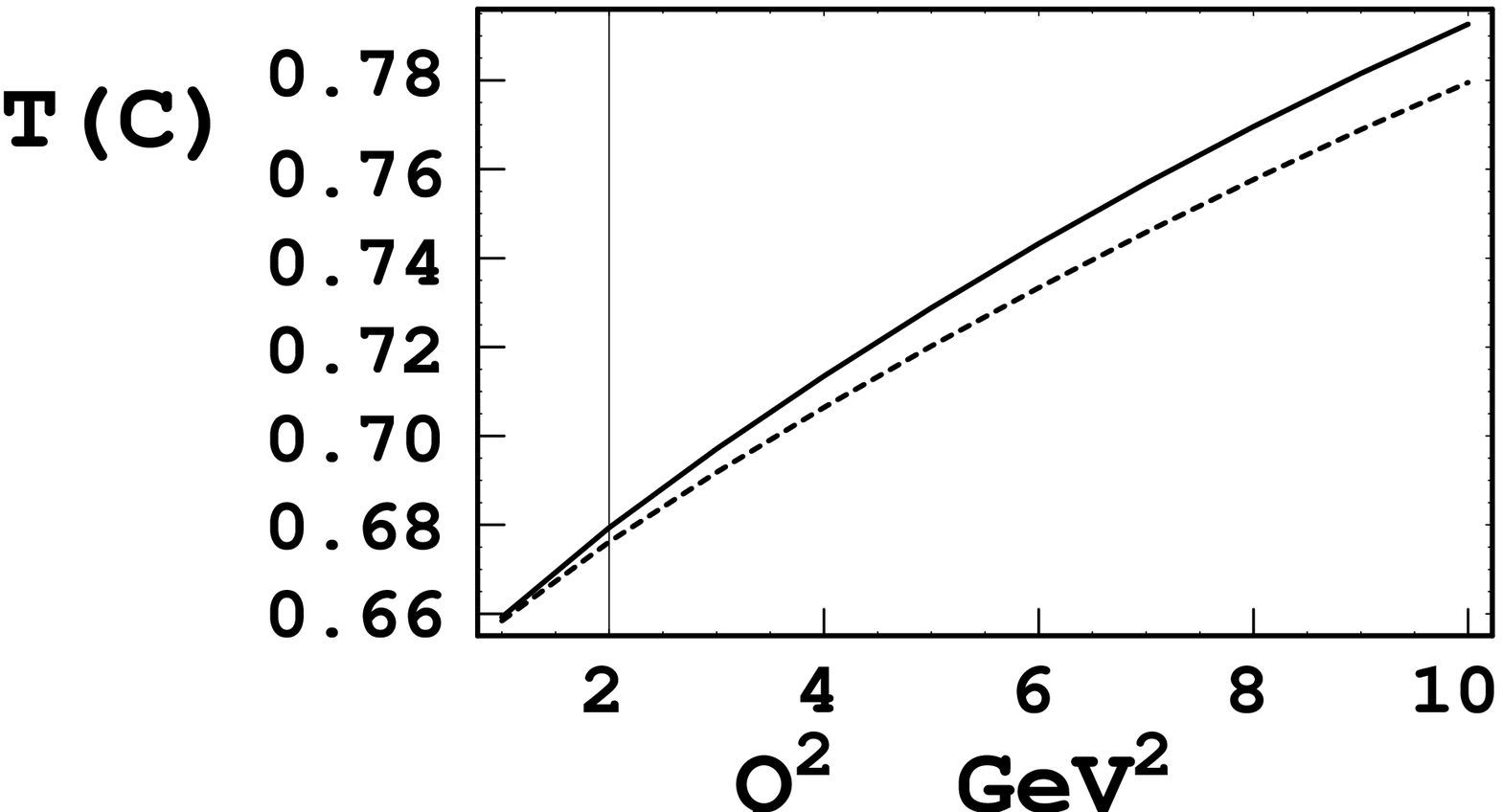}
\end{picture}
\caption{$^{12}$C Color transparency according to  \eq{T0CT} and \eq{sigplc}. Here
 $\sigma_{\rm tot}=30$  mb and   $l_c={2\nu\over (M_\rho^2+Q^2)} $ 
= 0.85 fm. The upper solid      curve is computed using
the standard value of $\Delta M^2=$0.7 GeV$^2$, and the lower dashed curve is computed using
 $\Delta M^2=$0.8 GeV$^2$.  }
\label{C1278comp}\end{figure}

\begin{figure}
\unitlength1.cm
\begin{picture}(14,8.2)(1.5,-.6)
\includegraphics{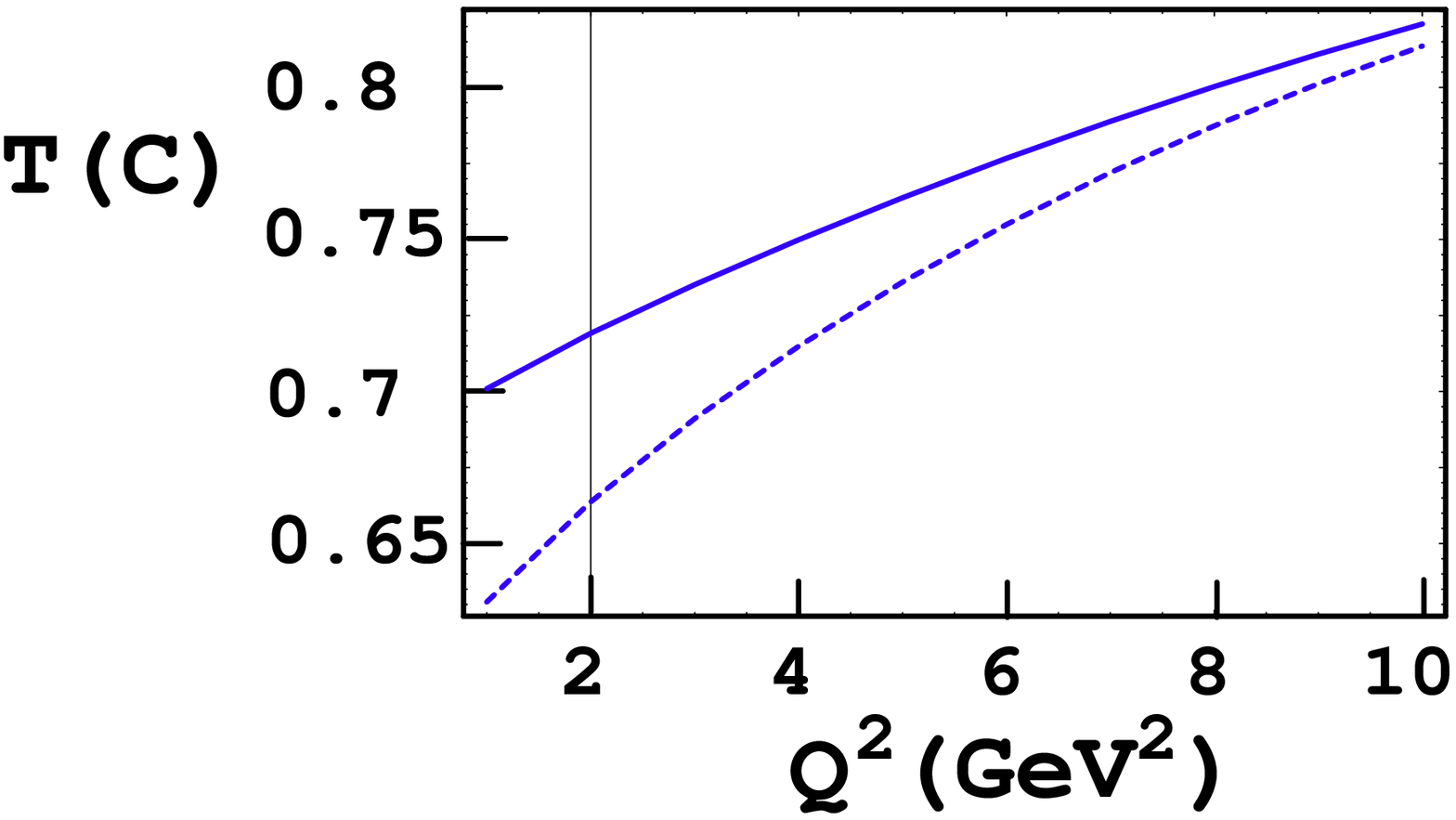}
\end{picture}
\caption{(Color online)  Effects of $\rho$ meson decay on  $^{12}C$. 
The parameters $l_c= $ 
 0.45 fm, and  $\sigma_{\rm tot}=25$  mb.  
The upper solid  curve is computed using the effects of
color transparency using $T_0$ of \eq{T0CTD} and \eq{sigplc}.
The lower dashed      curve represents color transparency as modified by 
 the use of \eq{sigplcdec} instead of  \eq{sigplc}.
}\label{c12ctdec2545}\end{figure}

\begin{figure}
\unitlength1.cm
\begin{picture}(14,8.2)(1.5,-.6)
\includegraphics{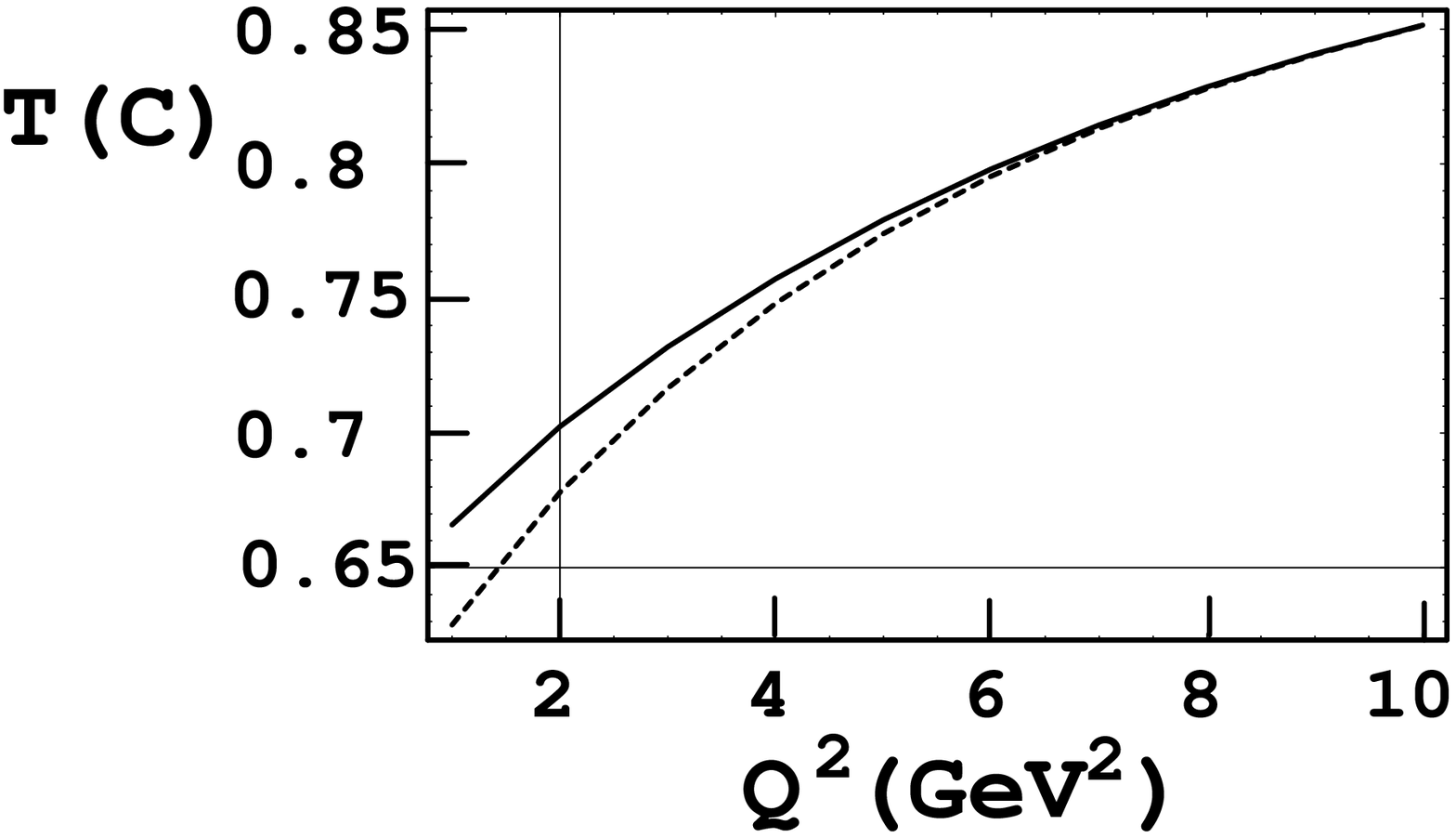}
\end{picture}
\caption{  Effects of $\rho$ meson decay on  $^{12}C$. 
The parameters $l_c= $ 
 0.85 fm, and  $\sigma_{\rm tot}=30$  mb.  
The upper solid   curve is computed using the effects of
color transparency using $T_0$ of \eq{T0CT} and \eq{sigplc}.
The lower dashed        curve represents color transparency as modified by 
 the use of \eq{sigplcdec} instead of  \eq{sigplc}.
}\label{c12ctdec32545}\end{figure}

Figure~\ref{c12ct25} shows the effects of color transparency as a function of $Q^2$
on a $^{12}$C target using
a $\rho-N$ cross section of 25 mb. Results for two different values of the coherence length
$l_c={2\nu\over (M_\rho^2+Q^2)} $, which determines $\nu$ and the rho meson momentum,
are shown. Figure~\ref{c12ct3} shows the effects of color transparency 
on a $^{12}$C target using
a $\rho-N$ cross section of 30 mb. We see that the effects of color transparency are expected
to be substantial if one is able to measure the cross section at large enough values of
$Q^2$. Increasing the value of $l_c$ for a fixed value of $Q^2$ increases the rho meson
momentum and therefore increases the  transparency. Figure~\ref{C1278comp} shows the
effects of increasing the value of $\Delta M^2$ from 0.7 to 0.8 GeV$^2$. There is only a negligible
effect, except for $Q^2\approx10$ GeV$^2$. Figs.~\ref{c12ctdec2545} and \ref{c12ctdec32545}
display the effects of including the effects of rho meson decay. These can be substantial
at low values of $Q^2$ if the $\rho N$ cross section is not too large.





\begin{figure}
\unitlength1.cm
\begin{picture}(14,8.2)(1.5,-.6)
\includegraphics{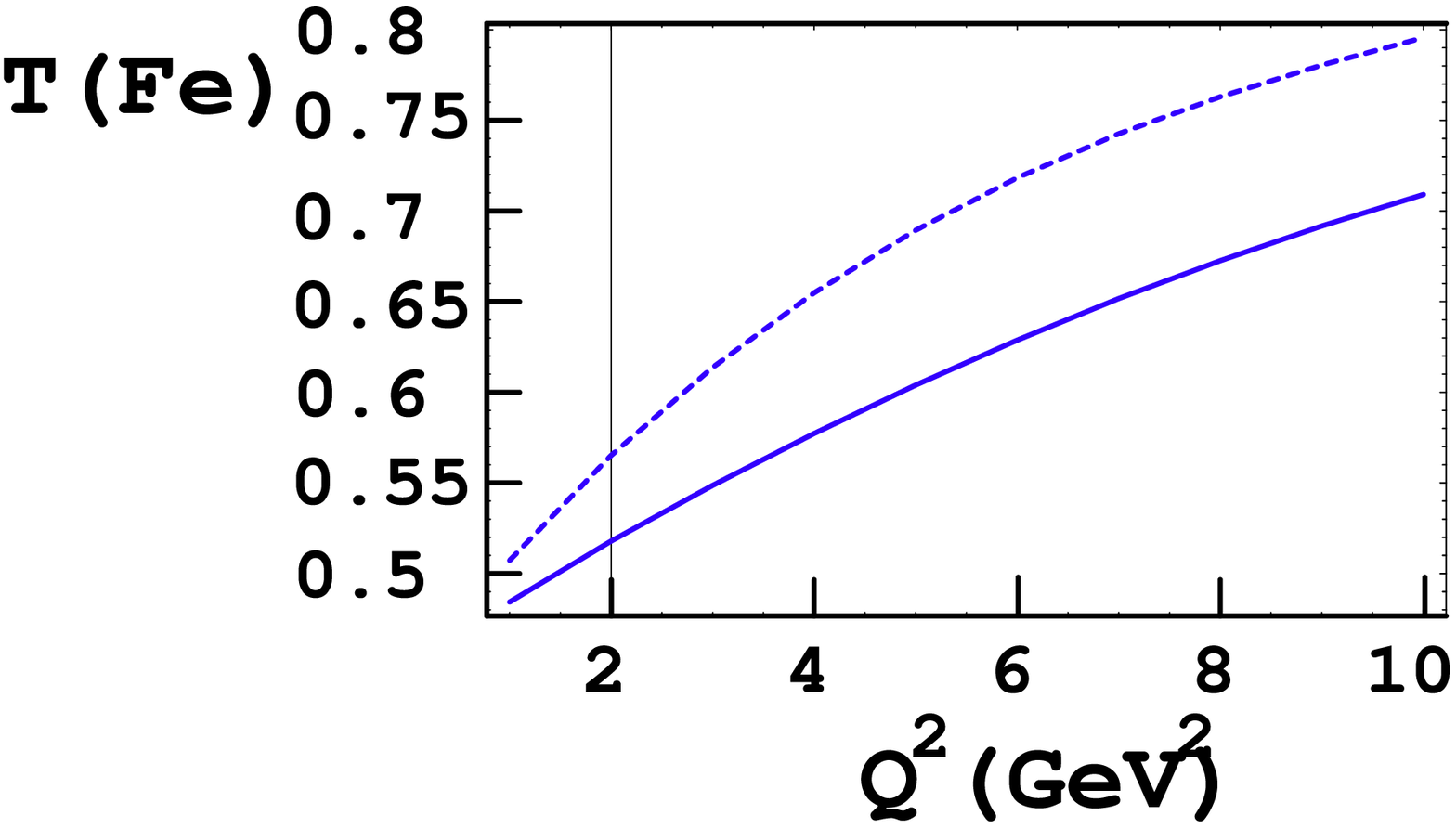}
\end{picture} 
\caption{(Color online) $^{56}Fe$ Color transparency using $T_0$ of \eq{T0CT}  and \eq{sigplc}
with $\sigma_{\rm tot}=25$  mb. The upper dashed curve is computed using $l_c={2\nu\over (M_\rho^2+Q^2)} $ 
= 0.85 fm.The lower solid  curve is computed using $l_c= $ 
 0.45 fm. 
 }%
\label{CompFe254585.eps}\end{figure}

\begin{figure}
\unitlength1.cm
\begin{picture}(14,8.2)(1.5,-.6)
\includegraphics{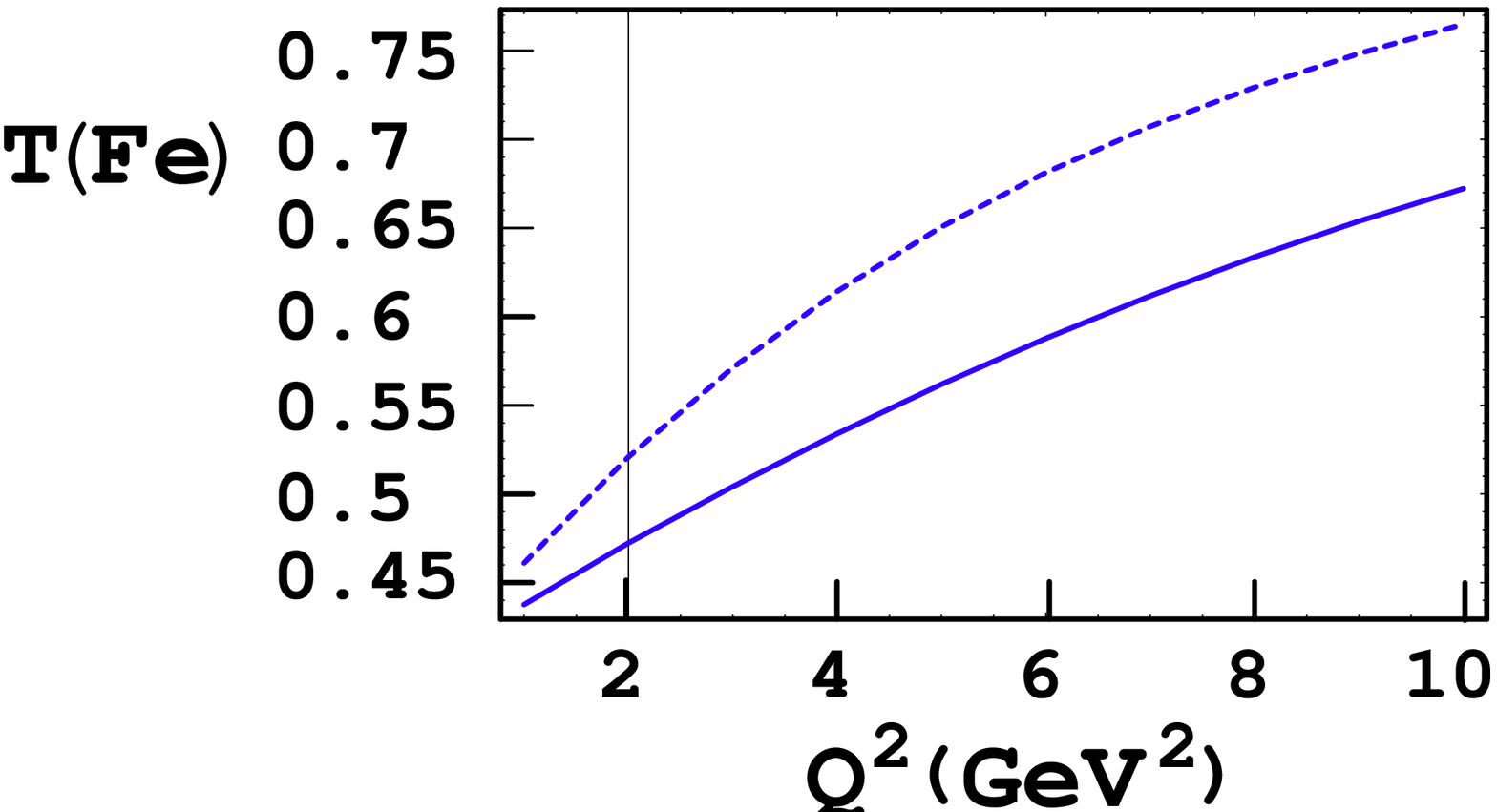}
\end{picture} 
\caption{(Color online) $^{56}Fe$ Color transparency using $T_0$ of \eq{T0CT}  and \eq{sigplc}
with $\sigma_{\rm tot}=30$  mb. The upper dashed curve is computed using $l_c={2\nu\over (M_\rho^2+Q^2)} $ 
= 0.85 fm.The lower solid  curve is computed using $l_c= $ 
 0.45 fm.
 }%
\label{CompFe304585.eps}\end{figure}

\begin{figure}
\unitlength1.cm
\begin{picture}(14,8.2)(1.5,-.6)
\includegraphics{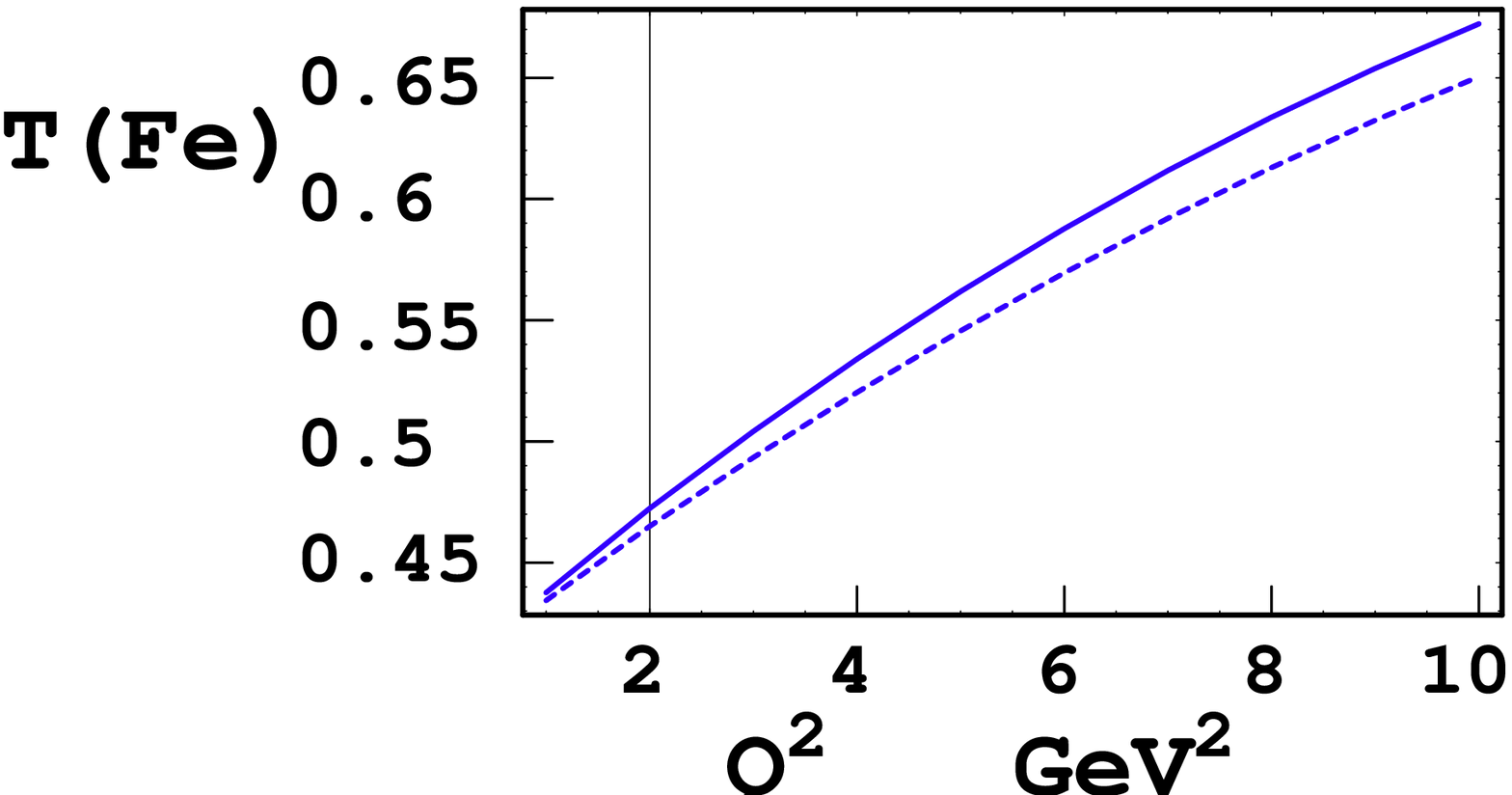}
\end{picture}
\caption{(Color online)$^{56}$Fe color transparency according to  \eq{T0CT} and \eq{sigplc}. Here
 $\sigma_{\rm tot}=30$  mb and   $l_c={2\nu\over (M_\rho^2+Q^2)} $ 
= 0.85 fm.. The upper solid      curve is computed using
the standard value of $\Delta M^2=$0.7 GeV$^2$, and the lower dashed curve is computed using
 $\Delta M^2=$0.8 GeV$^2$.  }
\label{Fe78comp}\end{figure}

\begin{figure}
\unitlength1.cm
\begin{picture}(14,8.2)(1.5,-.6)
\includegraphics{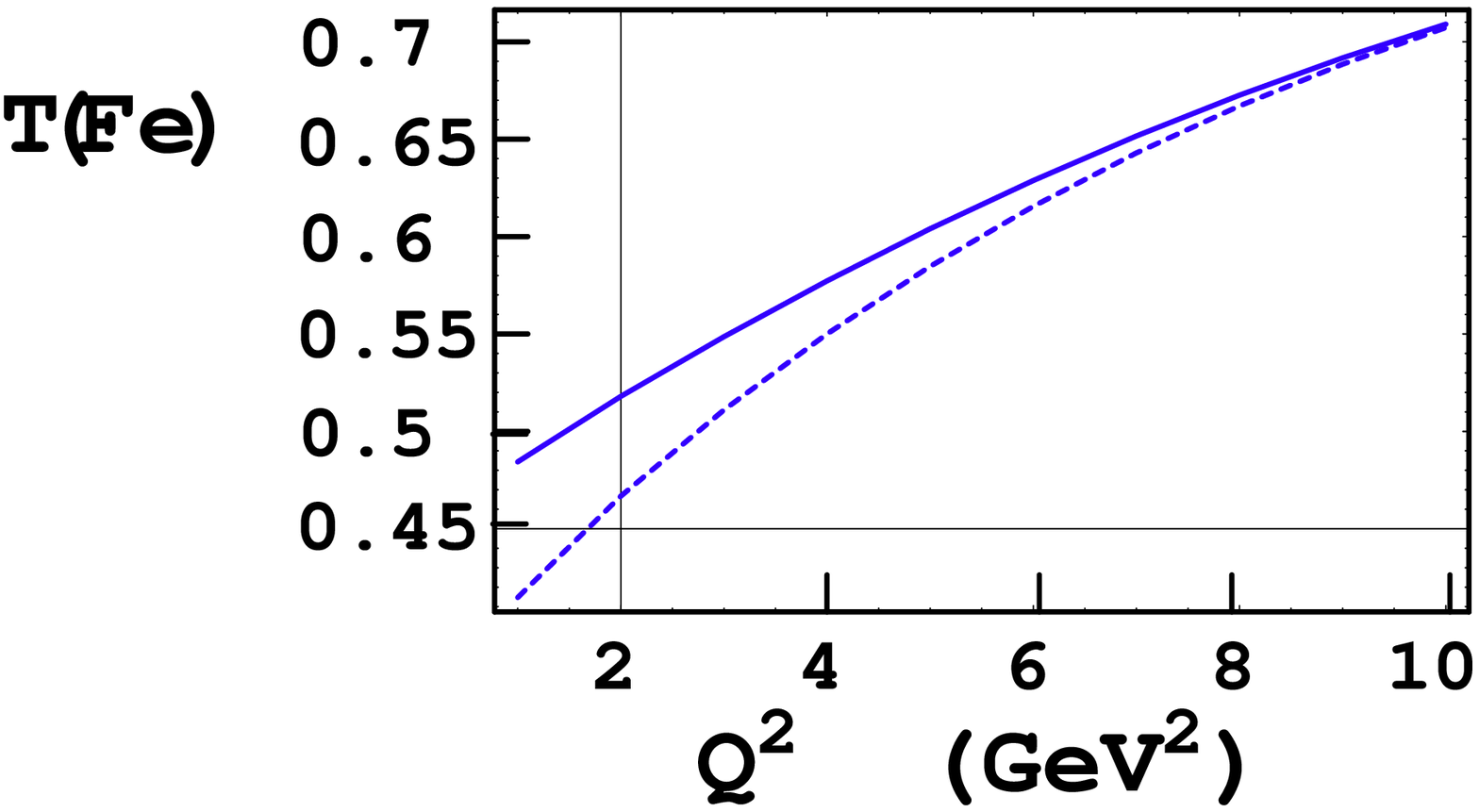}
\end{picture} 
\caption{(Color online) Effects of $\rho $ meson decay on
$^{56}Fe$ $T_0$ using 
$\sigma_{\rm tot}=25$  mb and  $l_c=$ 0.45 fm. The upper solid  curve is computed 
using   \eq{T0CT}  and \eq{sigplc} and the lower dashed curve  is computed using using   \eq{T0CT}  and \eq{sigplcdec} to include the effects of decay.
 }
\label{CompFedec2545.eps}\end{figure}

\begin{figure}
\unitlength1.cm
\begin{picture}(14,8.2)(1.5,-.6)
\includegraphics{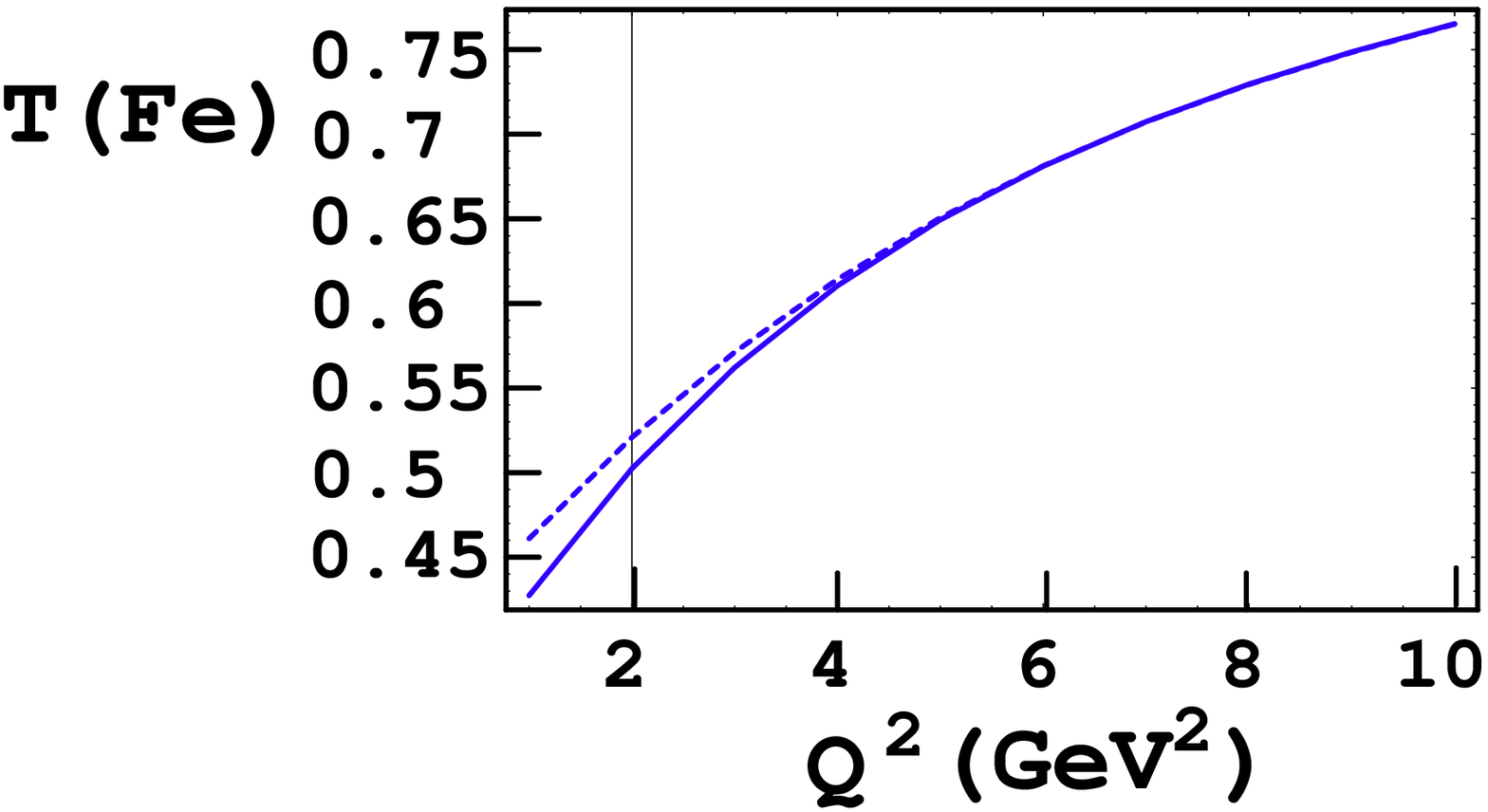}
\end{picture} 
\caption{(Color online) Effects of $\rho $ meson decay on
$^{56}Fe$ $T_0$ using 
$\sigma_{\rm tot}=30$  mb and  $l_c=$ 0.45 fm. The upper dashed   curve is computed 
using   \eq{T0CT}  and \eq{sigplc} and the lower solid  curve  is computed using using   \eq{T0CT}  and \eq{sigplcdec} to include the effects of decay.
 }
\label{CompFedec3085.eps}\end{figure}

Figure~\ref{CompFe254585.eps} shows the effects of color transparency as a function of $Q^2$
on a $^{56}$Fe target using
a $\rho-N$ cross section of 25 mb. Results for two different values of the coherence length
$l_c={2\nu\over (M_\rho^2+Q^2)} $, which determines $\nu$ and the rho meson momentum,
are shown, and there is a substantial difference.
 Figure~\ref{CompFe304585.eps} shows the effects of color transparency 
on a $^{56}$Fe target using
a $\rho-N$ cross section of 30 mb. Again, we see that the effects of color transparency are expected
to be substantial if one is able to measure the cross section at large enough values of
$Q^2$. Increasing the value of $l_c$ for a fixed value of $Q^2$ increases the rho meson
momentum and therefore increases the  transparency. Figure~\ref{Fe78comp} shows the
effects of increasing the value of $\Delta M^2$ from 0.7 to 0.8 GeV$^2$. Again there is only a negligible
effect, except for $Q^2\approx10$ GeV$^2$.   
Figs.~\ref{CompFedec2545.eps} and \ref{CompFedec3085.eps}
display the effects of including the effects of rho meson decay. These can be substantial
at low values of $Q^2$ if the $\rho N$ cross section is less than about 30 mb.


\section{Summary}
\label{sec:sum}
This work provides a reaction theory for electroproduction of vector  mesons, for 
$-t\gg 1/R_A^2$, that allows
one to assess the energy loss of each term in the multiple scattering series. 
These values of $-t$ are sufficient to justify our semi-classical picture.
We find
however, that the leading term $T_0$ of \eq{T0} and \eq{T0CT}, \eq{T0CTD} dominate for nuclei up
to Fe, as long as the momentum transfer $-t$ to the nucleus is sufficiently 
small.  Higher-order terms in the
multiple scattering series become important, causing a significant  
increase in the transparency
 as the value of $-t$ increases. This effect, entering in Glauber theory, could mimic a signal
of color transparency if the experimental
 acceptance in $-t$ increases with the virtuality $Q^2$ of the photon.
We also study the 
 effects  of $\rho$ meson decay inside the nucleus. These  are typically
about 5\%, as long as the momentum transfer $-t$ to the nucleus is small.
 This is large enough to influence the 
interpretation of the onset of color transparency. In 
particular, this effect disappears rapidly
as $Q^2$ increases from about 1 to 3 GeV$^2$, causing a rise in the transparency that can not
be interpreted as being related to color transparency.

We study the size of color transparency effects for C and Fe nuclei for values of $Q^2$ up to
10 GeV$^2$. 
The detailed results depend strongly on the assumed value of the $\rho N$ cross section.

The overall effects of color transparency are large for both nuclear targets if
$Q^2$ is greater than about 5 GeV$^2$. Testing our current theory, using the data of 
Ref.\cite{ANL} will  depend on the experimental
ability to push the combined statistical and systematic errors for $Q^2\le 3$ GeV$^2$ to
 better than about 10\%.  

\section*{Acknowledgments}
 This research was supported by 
the the US Israeli BNSF (LF) and by  the
United States Department
of Energy (GAM and MS). We thank K. Hafidi and  L. El Fassi for useful discussions regarding their experiment.

\end{document}